\documentclass[aps, prb, twocolumn,groupedaddress]{revtex4}

\usepackage{graphicx}
\usepackage{dcolumn}
\usepackage{bm}
\usepackage{ifthen}
\usepackage{psfrag}
\usepackage{booktabs}

\newcommand{\bb}{\begin{equation}}
\newcommand{\ee}{\end{equation}}
\newcommand{\ba}{\begin{eqnarray*}}
\newcommand{\ea}{\end{eqnarray*}}

\newcommand{\rr}{{\mathbf r}}
\newcommand{\dr}{{\rm d}{\bf r}}

\begin{document}

\title{Sedimentation of a two-dimensional colloidal mixture exhibiting liquid-liquid and gas-liquid phase separation: a dynamical density functional theory study}

\date{\today}

\author{Alexandr Malijevsk\'y}
\affiliation{E. H\'ala Laboratory of Thermodynamics, Institute of Chemical Process Fundamentals of ASCR, 16502 Prague 6, Czech Republic} \affiliation{Department of Physical Chemistry, Institute of Chemical Technology, Prague,
166 28 Praha 6, Czech Republic}

\author{Andrew J. Archer}
\affiliation{Department of Mathematical Sciences, Loughborough University, Loughborough LE11 3TU, UK}

\date{\today}

\begin{abstract}
We present dynamical density functional theory results for the time evolution of the density distribution of a sedimenting model two-dimensional binary mixture of colloids. The interplay between the bulk phase behaviour of the mixture, its interfacial properties at the confining walls, and the gravitational field gives rise to a rich variety of equilibrium and non-equilibrium morphologies. In the fluid state, the system exhibits both liquid-liquid and gas-liquid phase separation. As the system sediments, the phase separation significantly affects the dynamics and we explore situations where the final state is a coexistence of up to three different phases. Solving the dynamical equations in two-dimensions, we find that in certain situations the final density profiles of the two species have a symmetry that is different from that of the external potentials, which is perhaps surprising, given the statistical mechanics origin of the theory. The paper concludes with a discussion on this.
\end{abstract}

\maketitle

\section{Introduction}
Colloid science is a field where physical chemistry and statistical mechanics meet.  Colloidal suspensions exhibit a number of interesting phenomena, such as flocculation, adsorption (e.g.\ of polymers), bridging, osmosis,
Tyndall effect, etc. For applications, it is important to understand and control the stability of colloidal suspensions, i.e., to determine the conditions under which the colloids remain dispersed or when the colloids gather
to form clusters of a macroscopic size. The interactions between colloids are often dominated by excluded volume effects, and so statistical mechanical theories developed for simple fluids, based on a simple molecular model
description, can often be used to understand fundamental properties of colloidal systems. Even crude approximations, where only the repulsive forces between colloids are considered, can give good insight into the phase
behaviour of mixtures of colloids and the effects of packing, or `depletion' interactions, between the different species.\cite{BH} Clearly, more complex models are required for charged colloids that may be treated using DLVO
(Derjaguin--Landau--Verwey--Overbeek) theory, or for mixtures of colloids and (non-adsorbing) polymers, for which a simple Asakura--Oosawa--Vrij potential provides a surprisingly good description.\cite{BH} Colloidal fluids are
also interest because of the ability to observe directly the fluid phenomena which occur on length scales accessible to confocal microscopy and which are much harder to observe in molecular fluids. For example, in Refs.\
\onlinecite{ASL04, AL08} the detailed dynamics of fluid interfaces and the coalescence of liquid drops could be studied at the particle level.

There is also considerable interest in two-dimensional (2D) model colloidal fluids.  These enable the study of the properties of colloidal suspensions which are
effectively pinned to a 2D plane; this is the case for particles adhered by capillary forces to a fluid interface\cite{BrOe07} (either a liquid--gas or
liquid--liquid interface) or those whose motion in one dimension is restricted by an applied external field. For example, 2D models have been studied by means of
computer simulations \cite{sear,imperio} and density functional theory \cite{archer} to understand the origin of the structures formed by certain colloidal
monolayers at the air-water interface \cite{BrOe07,ghezzi}. In these particular systems one observes complex microphase-separated equilibrium structures of
various morphologies that form due to a competition between a short-range attraction and a longer-ranged repulsion between the colloids. Colloids at the
liquid--gas interface experience a capillary attraction between themselves, that can lead to aggregation and phase separation.\cite{BrOe07, oettel} Colloidal
monolayers that have sedimented onto the flat bottom of a container can also act as a 2D fluid, and binary colloidal mixtures of such particles can exhibit
freezing \cite{MBLB04} and phase separation\cite{zvyagolskaya}.

A description of the dynamical properties of such systems presents a challenging task.  Dynamical properties of fluids are commonly studied by solving the
Navier-Stokes equations that treat the fluid as a continuum. However, a more microscopic theory is required, if one wishes to obtain a description on the
colloidal length-scale. Over the last decade or so, a dynamical density functional theory (DDFT), originally formulated by Marconi and Tarazona,\cite{marconi,
MaTa00} has been developed to offer such a description for colloidal fluids. This approach stems from the assumption that the colloids can be modelled as an
assembly of Brownian particles whose motion is governed by stochastic (Langevin) equations of motion. By making the assumption that two-body correlations in a
fluid out of equilibrium can be approximated by those of an equilibrium fluid with the same one-body density profile, a deterministic equation of motion for the
one-body density $\rho(\rr, t)$ can be derived.\cite{marconi, MaTa00, ArEv04, ArRa04} The resulting equation of motion for the time evolution of $\rho(\rr, t)$
depends on the functional derivative of the equilibrium free energy functional, $F[\rho]$. Thus, the theory takes as input the accurate and verified
approximations for $F[\rho]$ that have been developed over the last 30 years or so in classical density functional theory (DFT).\cite{evans_hend} The advantage
of basing the dynamical theory on equilibrium DFT is that the approach guarantees (at the very least) a reliable prediction of the equilibrium fluid density
profile in the long time (equilibrium) limit, for external potentials that have no time dependence.

In our previous work in Ref.~\onlinecite{archer_mal}, we studied the interplay between sedimentation under the influence of a lateral driving force (e.g.\ due to gravity) and the gas-liquid phase separation of a 2D one-component colloidal suspension. Note that by colloidal `gas' we mean a low density suspension and by `liquid' we mean a high density suspension. This phase separation is driven by the attractive forces between the colloids and is completely analogous to the gas-liquid phase separation of molecular fluids. During this colloidal sedimentation the attractive forces between the colloids can lead to aggregation and the lateral driving also plays an important role in determining the final equilibrium state of the system and the degree to which the colloids are spread over the hard wall at the bottom of the system. The free energy functional used in Ref.\ \onlinecite{archer_mal} is a rather simple approximation, consisting of a local density approximation (LDA) for the hard core repulsive interactions between the colloids and a simple mean-field approximation for the attractive part of the pair interactions. Nevertheless, comparison of the DDFT results with Brownian dynamics simulations revealed surprisingly good agreement. However, this was only achieved when the DDFT was solved as a 2D problem allowing the system to exhibit, in some situations, symmetry breaking in the $x$-direction, which is the direction parallel to the wall (and perpendicular to the lateral driving force acting in the $y$-direction). When we also calculate the time evolution of the density profile assuming that it only varies in the $y$-direction (the direction in which the external potential varies), for cases where symmetry breaking is observed in the 2D DDFT calculation, we find that the agreement between the one-dimensional (1D) DDFT results with simulation is rather poor, with a significant overestimation of the average density in the vicinity of the wall. Much better results are obtained by averaging over the symmetry-broken density distribution obtained from the 2D DDFT -- i.e.\ averaging $\rho(x,y)$ over the $x$-direction. These 2D density profiles correspond to a series of distinct drops of the colloidal liquid phase. Such states of course only occur due to the fact that the system consists of a fixed number of colloids confined within a finite size system. The reasons for this is discussed in detail in Ref.~\onlinecite{archer_mal} and we also return to this matter in the last section of this paper.

In the present paper, we generalise our previous work\cite{archer_mal} by considering a 2D binary mixture. The good agreement between the simulation and DDFT
results for the one-component fluid gives us confidence that the DDFT also provides a reliable description of the mixture. Thus, here we solely present a DDFT
treatment of the binary fluid, going to systems with sufficient numbers of particles that simulation is prohibitive. The interplay between the lateral driving
(hereafter assumed to be due to a gravitational field and referred to as such), the (bulk) phase separation and ``surface'' (strictly, line, in 2D) effects, from
the interfaces that are produced between the different phases, is shown to produce an extremely rich behaviour. Our goal here is not to construct a global `phase
diagram' of all the possible morphologies that can be observed for this system, which would be an extremely lengthy task, even in the case when only a restricted
number of the parameters of the model are varied. Instead, we display some typical results which illustrate the complexity of the confined mixture, giving a
detailed discussion of the characteristic behaviour of selected systems. We find that, in some situations, small changes of the system composition can lead to
quite dramatic changes in the behaviour of the system. The model differs from the one-component fluid because the binary system has the possibility of exhibiting
three phase coexistence. Furthermore, not only is the bulk fluid phase behaviour of the mixture more complex than its one-component counterpart, but the dynamics
is also substantially richer.

This paper proceeds as follows: We start in Sec.\ \ref{sec:II} by defining our model system, specifying the nature of the interactions between the particles and
the external field. The DDFT equations are then briefly described, together with the simple approximation for the free energy functional that we use. In
Sec.~\ref{sec:IIC} we calculate the equilibrium bulk fluid phase diagram. In Sec.~\ref{sec:III} we present our 2D DDFT results. We first focus on the regime
where only liquid-liquid demixing is observed. Following this, we consider the regime where the system exhibits phase separation to form states with three-phase
coexistence. We conclude with a summary of our results and a discussion of some of the underlying theoretical issues.

\section{Theory and bulk properties}\label{sec:II}
\subsection{Model and equation of motion}

We consider a 2D model binary mixture of colloids that interact through the pair potentials
 \bb
 u_{ij}(r)=u_{ij}^{\rm hd}(r)+u_{ij}^{\rm att}(r)\,,\label{pot}
 \ee
where the index $i,j=1,2$ labels the two different species of colloids in the system. The pair potential is composed of two contributions, the purely repulsive
hard-disk potentials
  \bb
 u_{ij}^{\rm hd}(r)=\left\{\begin{array}{ll} \infty,&r\leq\sigma_{ij}\,,\\
0& r>\sigma_{ij}\end{array}\right.
 \ee
and we assume the following simple form for the attractive contribution to the pair potentials
 \bb
 u_{ij}^{\rm att}(r)=-\varepsilon_{ij}\exp\left(-r/\sigma_{ij}\right)\,. \label{eq:u_att}
 \ee
Here, $r=\sqrt{(x_a-x_b)^2+(y_a-y_b)^2}$ is the distance between a pair of particles (labelled with the indices $a$ and $b$), whose centers are at
$\rr_a=(x_a,y_a)$ and $\rr_b=(x_b,y_b)$. $\sigma_{ij}=(\sigma_{ii}+\sigma_{jj})/2$, where $\sigma_{ii}$ is a hard-core diameter of a species $i$, and the
parameter $\varepsilon_{ij}>0$ determines a strength of the attractive interactions. In total, there are $N=N_1+N_2$ particles in the system, where $N_1$ and
$N_2$ are the number of particles of species 1 and 2, respectively.

We consider the binary fluid when it is confined between two hard walls at $y=0$ and $y=L$ and also under the influence of a constant lateral driving force due
to gravity, with amplitude $|a_i|$. Thus, the external potential exerted on a particle of species $i$ is: \bb \varphi_{i}(x,y)=\varphi^{\rm
hw}_i(x,y)+\varphi^{\rm hw}_i(x,L-y)+a_iy \ee where the hard-wall potential \bb
\varphi^{\rm hw}_i(x,y)=\left\{\begin{array}{ll} \infty,&y\leq 0,\\
0,& y>0.\end{array}\right. \label{eq:hard_pot}
\ee
We define $a_i<0$ so that the colloids are pushed {\em down} to the lower wall at $y=0$. Note too that in our DDFT calculations we replace the hard wall potentials $\varphi^{\rm hw}_i$ by the following slightly softened potentials
 \bb
\beta \varphi_i^{\rm soft}(x,y)=\left\{\begin{array}{ll} \gamma \exp\left(-\left(\frac{y/\sigma_{ii}-\gamma}{\gamma}\right)^\gamma\right),&y\leq 0,\\
0,& y>0,\end{array}\right. \label{eq:soft_pot}
 \ee
with $\gamma=10$ and where $\beta=1/k_BT$ is the inverse temperature; $T$ is the temperature and $k_B$ is Boltzmann's constant. Using (\ref{eq:soft_pot}) rather
than (\ref{eq:hard_pot}) has almost no noticeable effect on the results, but it does make the numerical computations more stable.

We assume that the dynamics of the colloids is governed by the following over-damped stochastic equation of motion:
\bb
 \dot{\rr}_a= -\Gamma_i\nabla_a U(\{\rr_a\},t) + \Gamma_i{\bf X}_a(t),
 \label{eq:EOM}
\ee where ${\bf X}_a(t)$ is a white noise term, $\Gamma_i\equiv\beta D_i$ is the mobility, where $D_i$ is the diffusion coefficient for colloids of species $i$,
and \bb U(\{\rr_a\},t)=\sum_{a}^N\varphi_{i(a)}(\rr_a)+\sum_{a<b}u_{i(a)j(b)}(|\rr_a-\rr_b|) \ee with $i(a)=1$ for $a\le N_1$ and $i(a)=2$ for $a> N_1$, is the
potential energy of the system.\cite{Archer11}

\subsection{DFT and DDFT for the system}

DFT states that the equilibrium fluid density profiles of a binary mixture are those which minimise the grand potential functional:\cite{evans_hend}
 \bb
 \Omega[\rho_1(\rr),\rho_2(\rr)]=F[\rho_1(\rr),\rho_2(\rr)]-\sum_{i=1}^2\mu_i\int\dr\rho_i(\rr)\,,
 \ee
where $F$ is the Helmholtz free energy functional and $\mu_i$ is the chemical potential of a species $i$. The Helmholtz free energy functional can be written as
a sum of several contributions:
 \begin{eqnarray}
 &&F[\rho_1(\rr),\rho_2(\rr)]=k_BT\sum_{i=1}^2\int\dr \rho_i(\rr)\left[\ln[\Lambda_i^2\rho_i(\rr)]-1\right]\nonumber\\
&&+F_{\rm ex}[\rho_1(\rr),\rho_2(\rr)]+\sum_{i=1}^2\int\dr \rho_i(\rr)\varphi_i(\rr).\label{dft}
 \end{eqnarray}
The first term is the ideal-gas contribution, where $\Lambda_i$ is the thermal de Broglie wavelength of species $i$. The second term $F_{\rm ex}$ is the excess
contribution, due to the interactions between the colloids and the final term is the contribution due to the external potentials. The excess free energy can be
further split into two distinct contributions: The first from the repulsive part of the interactions between the particles and the second from the attractive
part, i.e.\ from the two different terms in the right hand side of (\ref{pot}), respectively:
 \bb
 F_{\rm ex}[\rho_1(\rr),\rho_2(\rr)]=F_{\rm hd}[\rho_1(\rr),\rho_2(\rr)]+ F_{\rm att}[\rho_1(\rr),\rho_2(\rr)]\,.\label{F_ex}
 \ee
The second term, stemming from the attractive interactions between the colloids, is approximated in a simple mean-field manner:
\bb
 F_{\rm att}
 =\frac{1}{2}\sum_{i,j=1}^2\int\dr\int\dr'\rho_i(\rr)\rho_j(\rr') u_{ij}^{\rm
 att}(|\rr-\rr'|)\,.\label{F_att}
 \ee
For the first term in Eq.\ (\ref{F_ex}), accounting for the repulsive hard-disk part of the interactions, we adopt a simple local density approximation
 \bb
 F_{\rm hd}=\int\dr f_{\rm hd}(\rho_1(\rr),\rho_2(\rr))\,,\label{F_hd}
 \ee
where $f_{\rm hd}$ is a free energy density for a mixture of hard-disks with uniform one-body densities $\{\rho_i\}$. Here, we use the approximation given by
scaled particle theory \cite{spt}
 \bb
f_{\rm hd}=k_BT\rho\left(-\ln(1-\eta)+\frac{\eta}{1-\eta}\right)\,,\label{f_hd}
 \ee
where the packing fraction $\eta=\frac{\pi}{4}\sum_i\rho_i\sigma_{ii}^2\equiv\sum_i\eta_i$, is the sum of the partial packing fractions, and the total density
$\rho=\rho_1+\rho_2$ is the sum of the densities of the two species.

Instead of the LDA approximation in Eq.\ (\ref{F_hd}), an improvement would be to use the non-local fundamental measure theory for hard disks, developed in Ref.\
\onlinecite{RMO12}. However, for the present system, due to the fact that the confining walls are hard, the LDA is surprisingly good.\cite{archer_mal} This is
because the density near to the wall is rather low and so the density profiles have very little of the oscillatory structure due to packing effects that fluids
normally have close to a wall to which the particles are attracted. The LDA is unable to describe such density oscillations, but since the amplitude of the
oscillations is very small in the present system, the LDA actually performs reasonably well.\cite{archer_mal}

We now consider the statistical mechanics of the system when it is out of equilibrium. The dynamics of the system can be described using DDFT. The quantity of
interest here is the non-equilibrium fluid one-body density profiles $\rho_i(\rr,t)$. This is defined as an average over all possible realisations of the white
noise in Eq.\ (\ref{eq:EOM}), starting from an ensemble of particle positions consistent with the initial system set-up. In the present case, we consider a
system where the particles are initially uniformly distributed between the two walls. On integrating the Smoluchowski equation corresponding to Eq.\
(\ref{eq:EOM}), we obtain the following pair of coupled equations for the time evolution of the one-body densities:\cite{marconi, MaTa00, ArEv04, ArRa04,
Archer11}
 \bb
  \frac{\partial\rho_i(\rr,t)}{\partial t}=\Gamma_i\nabla\cdot\left[\rho_i(\rr,t)\nabla\frac{\delta F[\rho_1(\rr),\rho_2(\rr)]}{\delta
  \rho_i(\rr,t)}\right]\,,\;\;\;i=1,2\,. \label{ddft}
 \ee
In deriving Eq.\ (\ref{ddft}) the assumption is made that the instantaneous two-body correlations can be approximated by those corresponding to an equilibrium
system with the same one-body density distributions. It is this assumption that leads to the {\em equilibrium} Helmholtz free energy functional being input into
Eq.\ (\ref{ddft}). Additionally, here we approximate $F[\{\rho_i(\rr)\}]$ using the functional given in Eqs.\ (\ref{dft})--(\ref{F_att}).

\subsection{Bulk fluid phase behaviour}\label{sec:IIC}

In the absence of any external potentials, the fluid densities are uniform and so the free energy density of our model fluid becomes:
\begin{eqnarray}
f=\frac{F}{V}&=&k_BT\sum_{i=1}^2\rho_i(\ln\Lambda_i^2\rho_i-1)+\frac{1}{2}\sum_{i,j=1}^2\rho_i\rho_j\alpha_{ij}\nonumber\\
 &&+k_BT\rho\left(-\ln(1-\eta)+\frac{\eta}{1-\eta}\right)\,,\label{F_bulk}
\end{eqnarray}
where $V$ is the area of the 2D system and $\alpha_{ij}\equiv\int u^{\rm att}_{ij}(\rr)\dr=-2\pi\varepsilon_{ij}\sigma_{ij}^2$.

For two bulk phases $A$ and $B$ to coexist, the temperature, chemical potentials and pressure in the two phases must be identical, i.e.:
\begin{eqnarray}
T(\{\rho_i^A\}) &=& T(\{\rho_i^B\})\nonumber\\
\mu_1(\{\rho_i^A\}) &=& \mu_1(\{\rho_i^B\})\nonumber\\
\mu_2(\{\rho_i^A\}) &=& \mu_2(\{\rho_i^B\})\nonumber\\
P(\{\rho_i^A\}) &=& P(\{\rho_i^B\})\label{eq:coex_cond}
\end{eqnarray}
In the present system, the chemical potentials are given by:
 \begin{eqnarray}
\mu_i=k_BT\left[\ln\frac{\rho_i}{1-\eta}+\frac{3\eta-2\eta^2}{(1-\eta)^2}\right]+\sum_{j=1}^2\rho_j\alpha_{ij}\, ,
 \end{eqnarray}
 and the pressure is
 \bb
P=\frac{k_BT\rho}{(1-\eta)^2}+\frac{1}{2}\sum_{i,j=1}^2\rho_i\rho_j\alpha_{ij}\,.
 \ee

\begin{figure}[t]
\includegraphics[width=8.5cm]{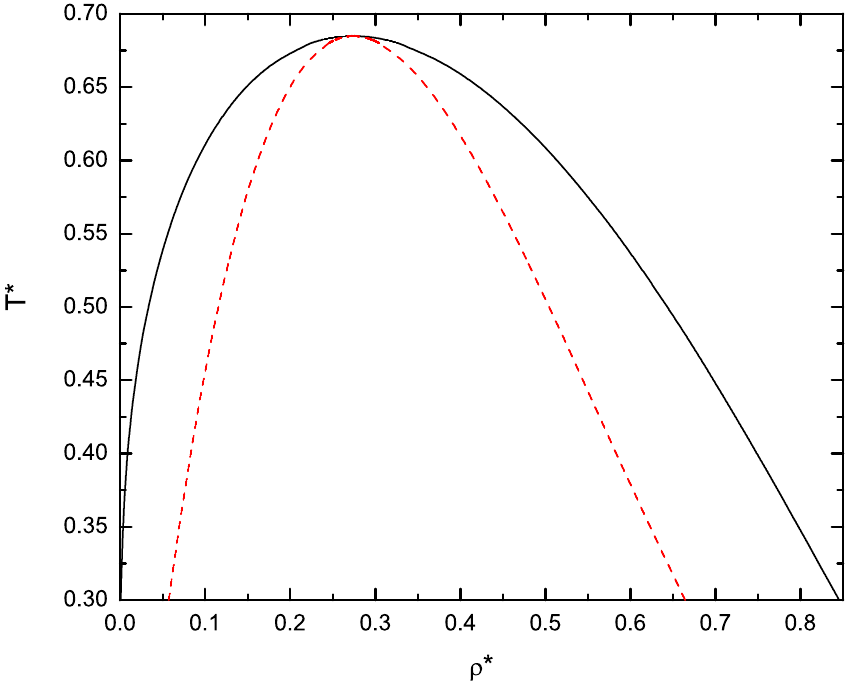}
\caption{The phase diagram of the one-component fluid in the dimensionless temperature $T^*=k_BT/\varepsilon$ versus density $\rho^*=\rho\sigma^2$ phase plane. The solid line is the gas-liquid coexistence (binodal) line and the dashed line is the spinodal.}
\label{fig_1c}
\end{figure}

\begin{figure}[t]
\includegraphics[width=9.cm]{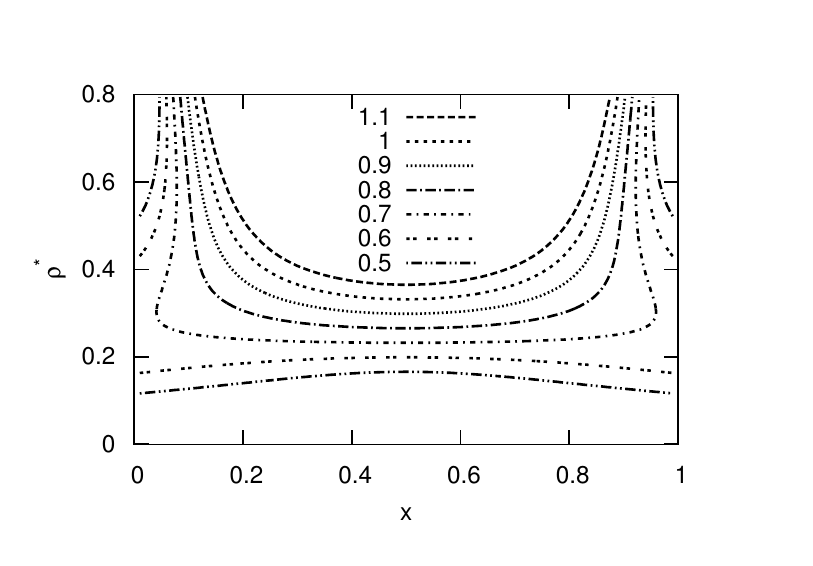}
\includegraphics[width=9.cm]{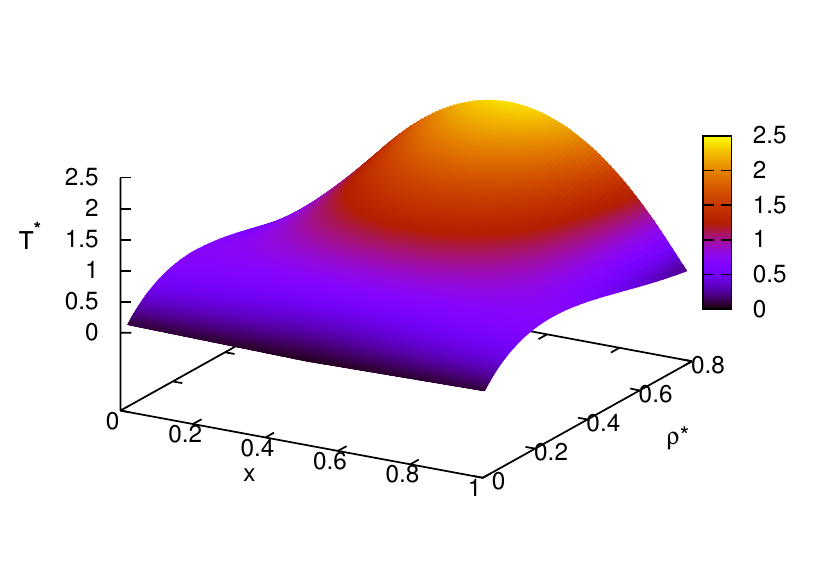}
\caption{Temperature at the limit of linear stability (spinodal) for a symmetric binary mixture, with the cross-interaction parameter $\varepsilon_{12}/\varepsilon=0.04$, as a function of the dimensionless total density $\rho^*$ and concentration $x$. In the upper figure we display contours of the spinodal surface, for various temperatures, given in the key. In the lower figure we display the spinodal as a surface plot of temperature $T^*$ as a function of $\rho^*$ and $x$. }\label{spinodals}
\label{fig_2a}
\end{figure}

Before proceeding to the description of the phase behaviour of the binary mixture, it is useful to construct the phase diagram for the model involving only one component. The temperature versus density phase diagram for the one-component fluid is displayed in Fig.\ \ref{fig_1c}, where both binodal and spinodal are displayed. The binodal is the locus of the coexisting densities and the spinodal is given by the condition $\frac{\partial^2 f}{\partial \rho^2}=0$. It defines the limit at which the uniform density state is linearly unstable.\cite{ArEv04} The critical density $\rho_c\sigma^2=0.274$ and critical temperature $k_BT_c/\varepsilon=0.685$ follow from the additional condition $\left.\frac{\partial^3 f}{\partial \rho^3}\right|_{\rho_c, T_c}=0$. Thus, when the uniform bulk fluid is cooled below the critical temperature, the system can phase separate, exhibiting gas-liquid phase coexistence.

Returning now to the binary fluid, in this case the spinodal limit of linear-stability condition is:
 \bb
 \frac{\partial\mu_1}{\partial\rho_1}\frac{\partial\mu_2}{\partial\rho_2}-\left(\frac{\partial\mu_1}{\partial\rho_2}\right)^2=0.\label{spin}
 \ee
This equation defines a temperature surface, as a function of the dimensionless total density $\rho^*=\rho_1\sigma_{11}^2+\rho_2 \sigma_{22}^2$ and concentration
$x=\rho_1\sigma_{11}^2/\rho^*$, below which a uniform fluid is linearly-unstable and will phase separate, exhibiting either gas-liquid coexistence, which is an
extension of the phase transition exhibited by the one-component fluid already discussed above, or the system can also exhibit liquid-liquid phase separation or
even three phase gas-liquid-liquid phase coexistence. An example of this spinodal surface is displayed in Fig.\ \ref{spinodals}.

In the remainder of this paper, we consider the case of a symmetric binary mixture, with $\sigma_{11}=\sigma_{22}=\sigma_{12}\equiv\sigma$, with $\varepsilon_{11}=\varepsilon_{22}\equiv\varepsilon$ and with equal mobilities
$\Gamma_1=\Gamma_2=\Gamma$. Thus, the like-species interactions for both species are identical. However, the cross-interaction potential between colloids of species 1 with species 2 is not the same, because we choose
$\varepsilon_{12}\neq\varepsilon$, specifically, we set $\varepsilon_{12}/\varepsilon=0.04$. Thus, the present system is rather similar to the 2D binary colloidal fluid studied in Ref.\ \onlinecite{zvyagolskaya}. However, in
the present work we use the much simpler form for the attractive interactions between the colloids given in Eq.\ (\ref{eq:u_att}). Taking $\sigma$ and $\varepsilon$ as the length and energy units, respectively, we define
reduced quantities such as the dimensionless temperature $T^*=k_BT/\varepsilon$, density $\rho^*=\rho\sigma^2$ and time $t^*=(\Gamma k_BT/\sigma^2)t$, used throughout this paper. The spinodal surface for the system with
$\varepsilon_{12}/\varepsilon=0.04$ is displayed in Fig.\ \ref{spinodals}. We do not display the binodal surface obtained from solving Eqs.\ (\ref{eq:coex_cond}), which lies above the spinodal surface, touching it only along
lines of critical points. Slices through such a coexistence surface can be seen for example in Refs.\ \onlinecite{RoSw, Woywod,Archer36}.

\section{DDFT results}\label{sec:III}

\begin{figure*}[ht]
\includegraphics[width=1.\textwidth]{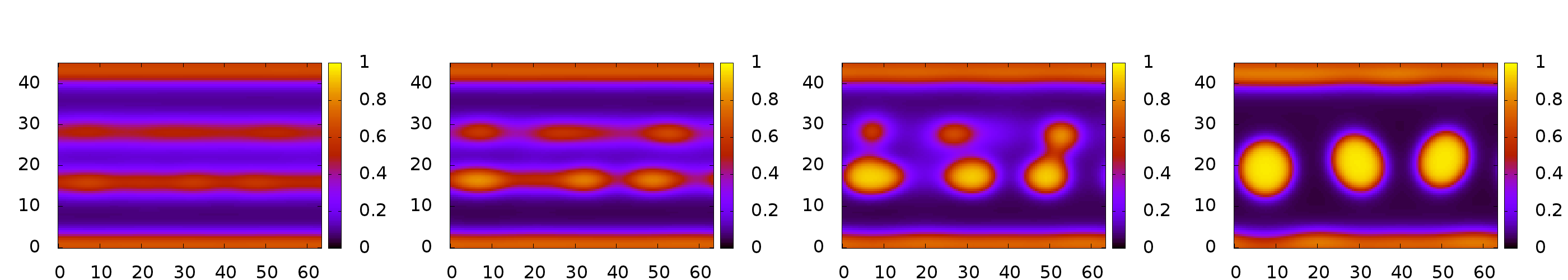}
\includegraphics[width=1.\textwidth]{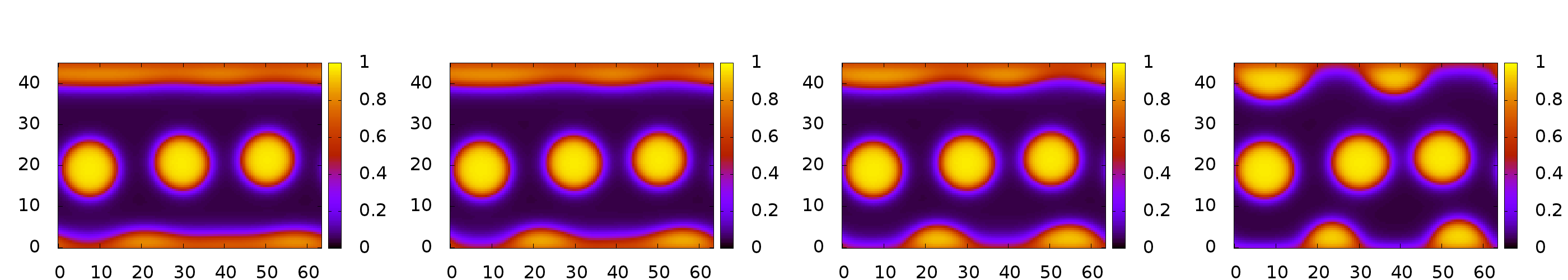}
\caption{A time series of local concentration profiles $c(\rr,t)$, for the case $a_1=a_2=0.01\varepsilon$, $\rho^*=0.6$, $T^*=1$, and $x=0.3$. The plots correspond to the following times elapsed after the system was initiated in a uniform mixed state (from top left to bottom right): $t^*=100, 150, 200, 300, 400, 500, 600$, and $1000$.} \label{conc_T1_rho6_a001_x03}
\end{figure*}

In this section we present typical results for the dynamical properties of the model as determined from DDFT calculations. Eqs.\ (\ref{ddft}) are solved by discretising the 2D system on a Cartesian grid of size
$64\,\sigma\times64\,\sigma$, with grid-spacing $\Delta x=\Delta y=0.5\sigma$. The two hard walls are located at $y=0$ and $y=L$ and separated by a distance $L=45\sigma$. There are periodic boundary conditions in the
$x$-direction, that runs parallel to the walls. We set the initial time $t=0$ state to correspond to that where the particles are uniformly mixed and uniformly distributed between the two confining walls. The corresponding
density profiles are $\rho_i(x,y)=\bar{\rho}_i+\delta_\rho(x,y)$ for $0<y<L$ and $\rho_i(x,y)=0$ otherwise, where $\bar{\rho}_i$ is the average fluid density of species $i$ in the system and $\delta_\rho$ is a small random
value, which is added as a perturbation to the density profiles in order to allow the system to form states that break the symmetry of the confining potentials; $\delta_\rho(x,y)$ is a randomly generated number from the
uniformly distributed interval $(-\bar{\rho}_i/20, \bar{\rho}_i/20)$. The gravitational force pushes the colloids towards the bottom confining wall at $y=0$. This is the same confining potential as was considered in our
earlier work in Ref.\ \onlinecite{archer_mal}, for the one-component fluid.

For the one-component fluid there are two main factors that determine the final state of the system:\cite{archer_mal} i) the strength of the external field pushing the particles towards the bottom of the system and ii) the strength of the interactions between the particles (in particular, whether it is strong enough to lead to phase separation). Since the confining walls are purely repulsive, the colloidal density right near the surface of the walls is low when there is attraction between the particles. This is because any given colloid that is close to the wall can lower its energy by moving away to be closer to the other colloids -- i.e.\ the colloidal fluid does not spread at the wall, unless the gravitational force pushing the particles downwards is strong enough to overcome the cohesive forces between the colloids, or the density of particles in the system is high enough that the weight of the colloids above or the upper wall pushes the lower colloids onto the bottom wall. If this is not the case and when the attraction between the colloids is sufficiently strong, then the colloids gather together to form a series of drops on the bottom wall, with a contact angle that can be close to 180$^\circ$.

In the present case of a binary mixture, much more complex behaviour is observed. This is, of course, partly due to a richer bulk fluid phase behaviour, compared to the one-component case -- see Sec.\ \ref{sec:IIC}. In addition, the effect of the external fields, both the gravitational component and the hard-walls, is particularly important and gives rise to a number of different morphologies of the fluid. Furthermore, as the system evolves it can form one morphology that then can become unstable after a certain time period and may then transform to another morphology before the equilibrium is reached. When the gravitational force is relatively weak, at early times, the main effects on the dynamics of the system are the thermodynamic forces driving the demixing of the fluid, coupled with the presence of the confining walls, which together lead to a surface-directed spinodal decomposition. This process is well understood in simple models and at early times in the dynamics, the behaviour of the present system is very similar to that observed in the more simple systems.\cite{BPF99, Puri05} However, the system we consider here also exhibits gas-liquid demixing (in addition to the liquid-liquid demixing) and this, together with the gravitational force pushing the particles towards the lower wall, lead to a very rich later stage dynamics.

\begin{figure*}[t]
\includegraphics[width=1.\textwidth]{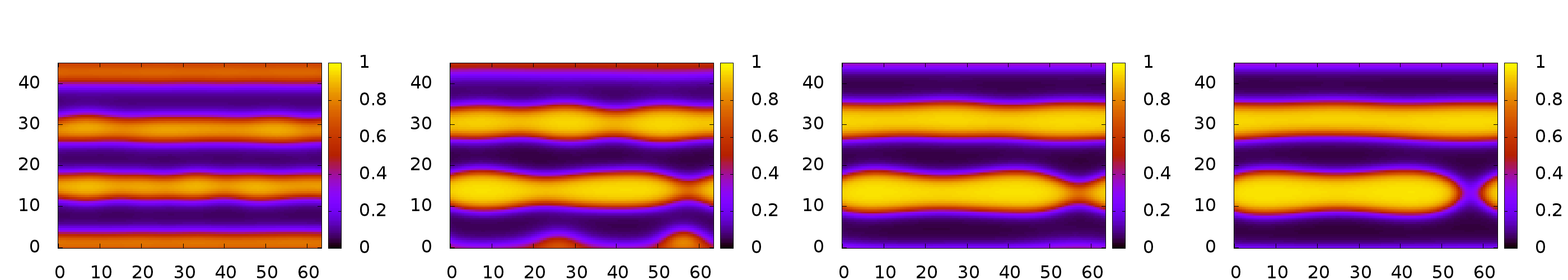}
\caption{A time series of local concentration profiles $c(\rr,t)$, for the case $a_1=a_2=0.01\varepsilon$, $\bar{\rho}^*=0.6$, $T^*=1$, and $x=0.4$. The plots correspond to the following times elapsed from the initial configuration (from
left to right): $t^*=100, 500, 700$, and $1000$.} \label{conc_T1_rho6_a001_x04}
\end{figure*}

\begin{figure*}[t]
\includegraphics[width=1.\textwidth]{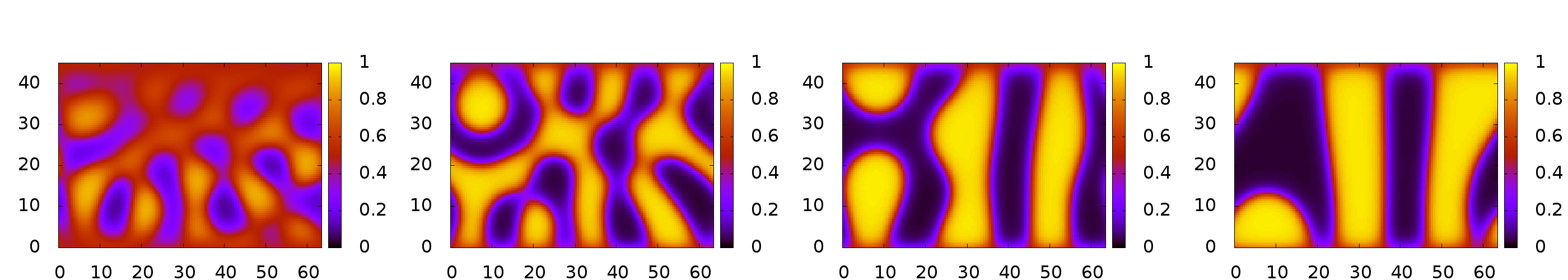}
\caption{A time series of local concentration profiles $c(\rr,t)$, for the case $a_1=a_2=0.01\varepsilon$, $\bar{\rho}^*=0.6$, $T^*=1$, and $x=0.5$. The plots correspond to the following times elapsed from the initial configuration (from
left to right): $t^*=100, 200, 500$, and $1600$.} \label{conc_T1_rho6_a001_x05}
\end{figure*}

\begin{figure*}[t]
\includegraphics[width=1.\textwidth]{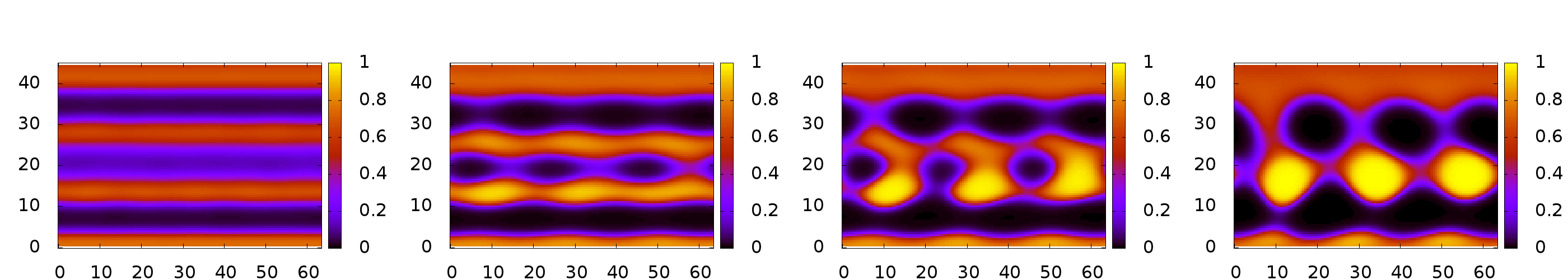}
\includegraphics[width=1.\textwidth]{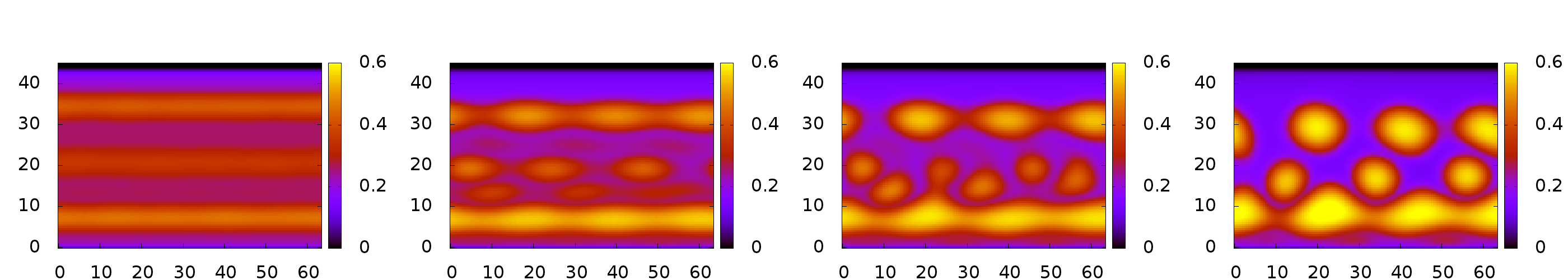}
\caption{A time series of local concentration profiles $c(\rr,t)$, (upper panels) and the corresponding total number density $\rho(\rr,t)\equiv\rho_1(\rr,t)+\rho_2(\rr,t)$ (lower panels) for $a_1=a_2=0.01\varepsilon$, $\bar{\rho}^*=0.3$, $T^*=0.5$, and $x=0.3$. The plots correspond to the following times (from left to right): $t^*=100, 400, 500$, and $600$.} \label{T05_rho03_a001_x03}
\end{figure*}

\begin{figure*}[t]
\includegraphics[width=1.\textwidth]{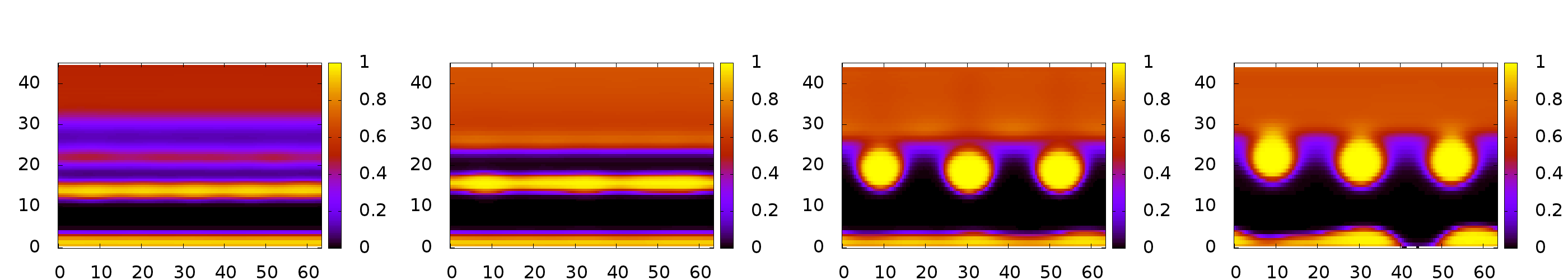}
\includegraphics[width=1.\textwidth]{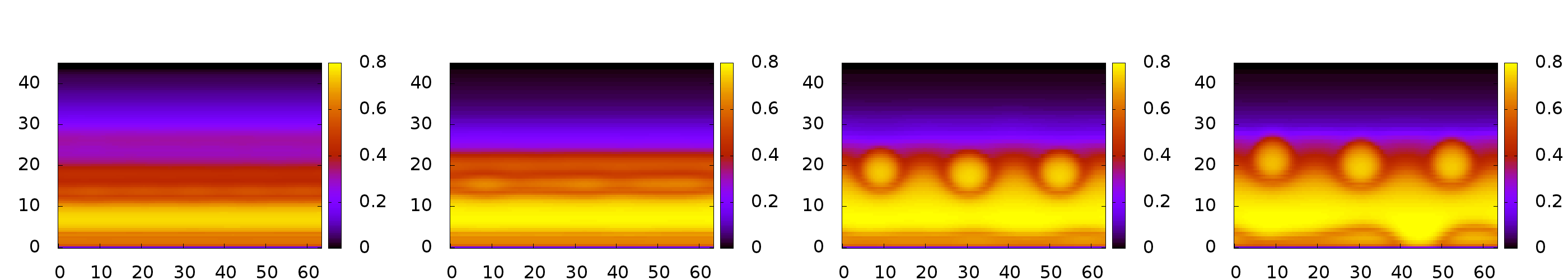}
\caption{A time series of local concentration profiles $c(\rr,t)$, (upper panels) and the corresponding total number density $\rho(\rr,t)\equiv\rho_1(\rr,t)+\rho_2(\rr,t)$ (lower panels) for $a_1=a_2=0.1\varepsilon$, $\bar{\rho}^*=0.3$, $T^*=0.5$, and $x=0.3$. The plots correspond to the following times (from left to right): $t^*=100, 200, 400$, and $600$.} \label{T05_rho03_a01_x03}
\end{figure*}

\begin{figure}[t]
\includegraphics[width=0.48\textwidth]{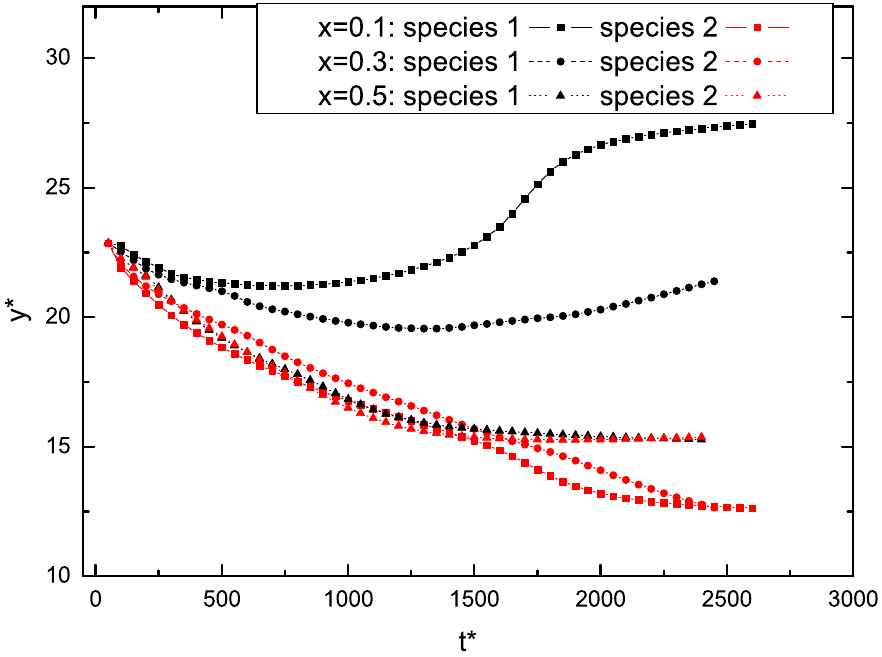}
\includegraphics[width=0.48\textwidth]{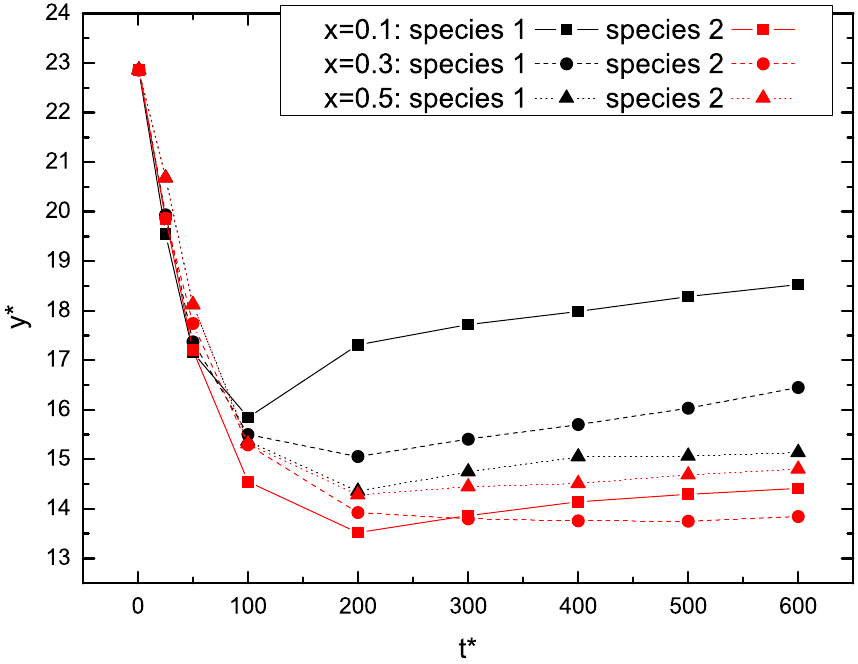}
\caption{Time-dependence of the height of the center-of-mass of both species for various different concentrations $x$, as given in the key, for $T^*=0.5$ and $\bar{\rho}^*=0.3$. Upper panel: for $a_1=a_2=0.01\varepsilon$, lower panel: for $a_1=a_2=0.1\varepsilon$.}
\label{y_rho03_T05}
\end{figure}

\begin{figure*}[t]
\includegraphics[width=1.\textwidth]{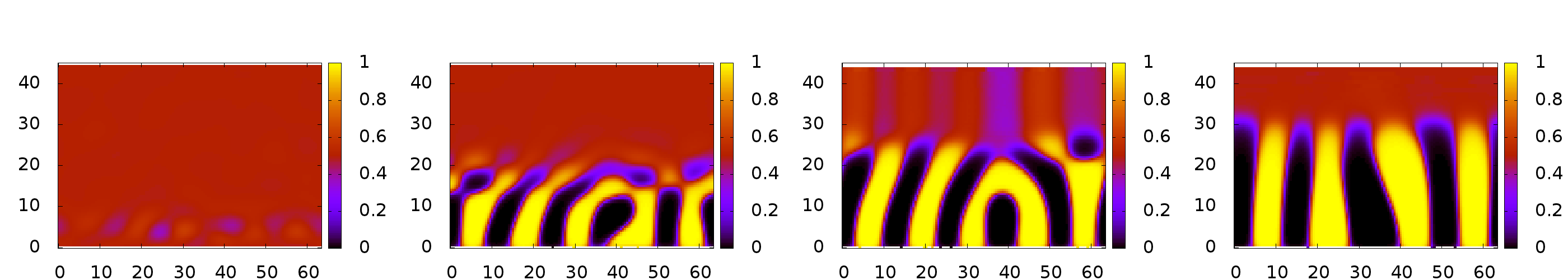}
\includegraphics[width=1.\textwidth]{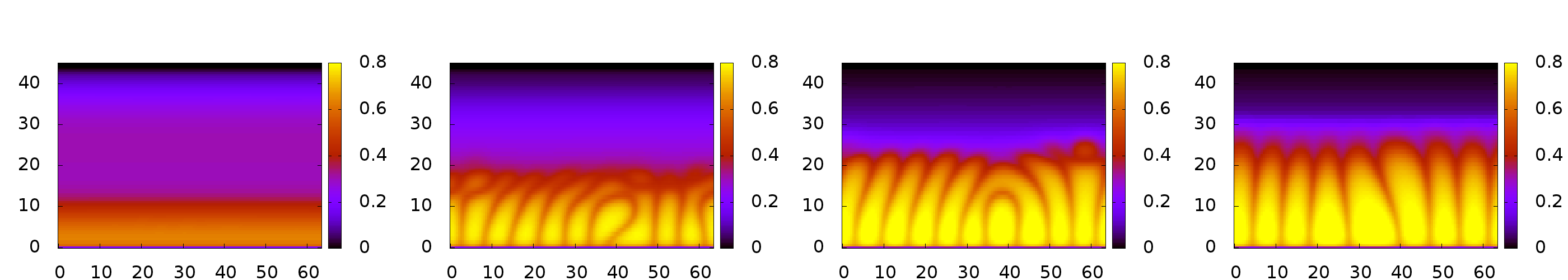}
\caption{A time series of local concentration profiles $c(\rr,t)$, (upper panels) and the corresponding total number density $\rho(\rr,t)\equiv\rho_1(\rr,t)+\rho_2(\rr,t)$ (lower panels) for $a_1=a_2=0.1\varepsilon$, $\bar{\rho}^*=0.3$, $T^*=0.5$, and $x=0.5$. The plots correspond to the following times (from left to right): $t^*=50, 100, 200$, and $600$.} \label{T05_rho03_a01_x05}
\end{figure*}

\begin{figure*}[t]
\includegraphics[width=1.\textwidth]{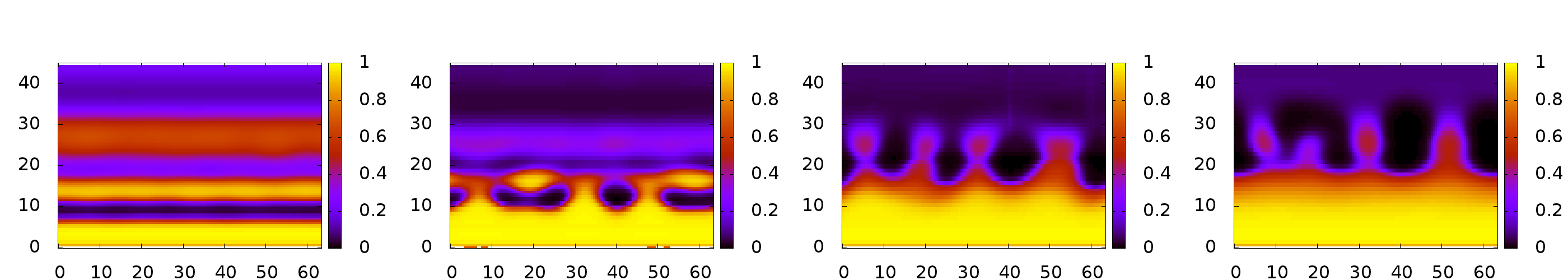}
\includegraphics[width=1.\textwidth]{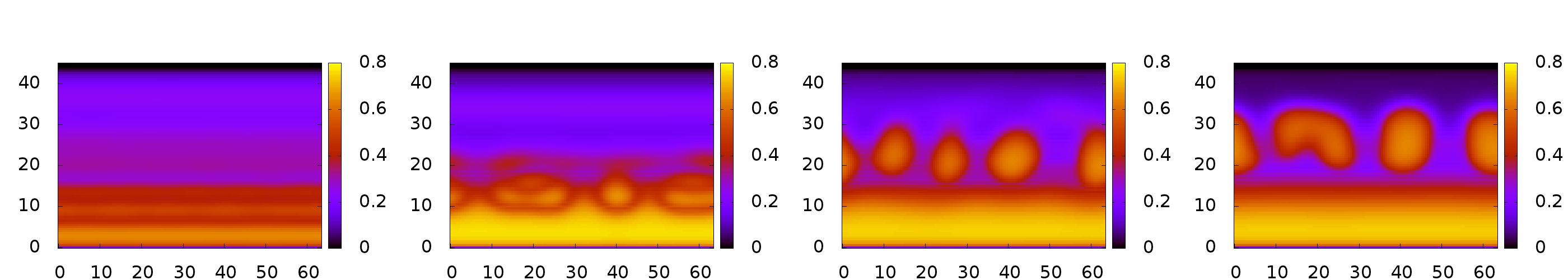}
\caption{A time series of local concentration profiles $c(\rr,t)$, (upper panels) and the corresponding total number density $\rho(\rr,t)\equiv\rho_1(\rr,t)+\rho_2(\rr,t)$ (lower panels) for $a_1=0.1\varepsilon$, $a_2=0.01\varepsilon$, $\bar{\rho}^*=0.3$, $T^*=0.5$, and
$x=0.5$. The plots correspond to the following times (from left to right): $t^*=100, 240, 400$, and $650$.} \label{T05_rho03_a001_a01_x05}
\end{figure*}

We start our survey of typical results with the case when both species are subject to an equally strong external field, $a_1=a_2=0.01\varepsilon$, and when the dimensionless temperature $T^*=1$ and average density in the system $\rho^*=0.6$, corresponding to a state point where there is no liquid-gas separation but liquid-liquid demixing is possible, depending on the concentration of the mixture. In Fig.\ \ref{conc_T1_rho6_a001_x03} we display a time series of plots of the local species 1 concentration, defined as follows: $c(\rr,t)=\frac{\rho_1(\rr,t)}{\rho_1(\rr,t)+\rho_2(\rr,t)}$. The total average concentration in the system $x=\frac{N_1}{(N_1+N_2)}$ is fixed. In Fig.\ \ref{conc_T1_rho6_a001_x03} we see in the early stages the strong influence of the walls and the system forms several alternating horizontal layers with the layers closest to the walls rich in the minority species 1. The demixing is due to the fact that the attraction between like species particles is much stronger than that between unlike species. The layering is parallel to the walls because the presence of the walls enhances the growth of demixing-density fluctuations perpendicular to the wall. However, due to the fact that there is a greater amount of species 2 in the system, such a configuration is not energetically favourable because the total amount of interface between the two demixed phases is fairly large in this configuration and so is the interfacial tension contribution to the free energy. The system can lower the free energy by reducing the amount of interface between the two demixed phases. Consequently, the thin layers rich in species 1 become eventually unstable with respect to the density fluctuations parallel to the walls and break-up into a number of droplets. It is interesting to note, however, that this instability hits only the inner layers while the layers at the top and at the bottom nearest to the walls survive long after the first droplets appear. This is because the walls have a stabilising effect on these layers. The distribution of radii of the drops emerging as the layers break-up is quite broad. The larger drops then grow at the expense of the smaller ones until the small ones vanish. Since the gravitational force is relatively weak and equal for both species in this case, the three resulting drops remain in the center of the system, surrounded by species 2 rich fluid. However, as soon as one of the drops (the one on the left in our plot) slightly shifts down, whether due to the external field or due to the randomness in the initial density profiles, a disturbance of both bottom and top layers becomes evident. This is because as the drop moves down, it pushes the fluid beneath which, in turn, pushes on the surface of the bottom layer. At the same time, an inverse process occurs above the drop: here the fluid is dragged in the direction of the moving drop which causes a local thickening of the top layer. Following this, over a rather long time period, both of the remaining stripes become unstable and break up into drops on the walls.

In Fig.\ \ref{conc_T1_rho6_a001_x04} we display results for a higher value of the total concentration, $x=0.4$. At early times, the system behaves similarly to in the previous case in Fig.\ \ref{conc_T1_rho6_a001_x03} (which was for $x=0.3$). Again, four layers rich of species 1 are formed, separated by three layers of liquid rich in species 2. However, in this case, the inner layers rich in species 1 are much thicker than previously, because of the increased amount of species 1 particles in the system. We clearly observe the growth of modulations along the borders of the inner stripes. However, there is the competing effect of a transfer of material from the outer stripes (adjacent to the walls) to the inner stripes, which results in the inner ones becoming `fatter', so that they remain intact over a long time period. Interestingly, the mechanism for the transfer of material from the outer stripes to the inner ones is via diffusion through the other liquid phase, rather than via the coalescence of drops, as was the case in Fig.\ \ref{conc_T1_rho6_a001_x03}. During this process the top layer continuously shrinks, but the layer at the bottom additionally becomes unstable and forms drops at the wall that then vanish over time in favour of the layer above. The remaining two thick layers seem to be quite stable, although breaks in the layer can and do form.

Next, we increase the amount of species 1, so that the concentration is $x=0.5$, i.e.\ there is exactly the same amount of each species of particles in the system. A time series of local concentration profiles is displayed in
Fig.\ \ref{conc_T1_rho6_a001_x05}. Owing to the fact that this is a symmetric mixture, the effect of the wall in directing the spinodal-decomposition by favouring the minority species is not present in this case. Instead, the
system initially phase separates away from the walls in the bulk of the fluid in an isotropic fashion. However, over time, the effect of the walls becomes evident in the orientation of the local concentration profiles and also
the interfaces between the two phases sharpen. There is also a striking difference from the previous case: Since in this case the mixture is completely symmetric and therefore neither species can be preferentially adsorbed at
the walls, the resulting morphology corresponds to vertical rather than horizontal stripes.

So far, we have considered systems where the temperature and density are both rather high and shown typical microstructure morphologies characteristic for systems exhibiting liquid-liquid demixing. It should be noted that in
these cases the effect of gravity is small and the evolution of these systems is very similar to the case with no gravity, $a_i=0$. For stronger fields, such as $a_1=a_2=0.1\varepsilon$ (not shown here), the shift downwards of
the center of mass of the entire system becomes substantial, leading to strong gradients of the density along the $y$-direction. The difference in densities at the top and bottom of the system can be as high as 50\%.
Nevertheless, apart from that, the sedimentation relaxation remains qualitatively similar to that presented above for the weaker field. We now show results corresponding to the lower temperature $T^*=0.5$, i.e.\ below the
critical temperature of the gas-liquid phase separation, c.f.\ Fig.\ \ref{fig_2a}. We also consider cases where the overall density in the system is lower, so that there is the possibility of observing the gas phase in
addition to the two liquid phases. In Fig.\ \ref{T05_rho03_a001_x03} we display a time series of local concentration profiles $c(\rr,t)$ and in addition the corresponding total density profiles
$\rho(\rr,t)\equiv\rho_1(\rr,t)+\rho_2(\rr,t)$ for a system with average overall density $\rho^*=0.3$, concentration $x=0.3$ and again subject to external potentials with weak gravitational driving towards the bottom of the
system, $a_1=a_2=0.01\varepsilon$. In the initial stages, the system evolves in a manner quite similar to what was observed in Figs.\ \ref{conc_T1_rho6_a001_x03} and \ref{conc_T1_rho6_a001_x04}, i.e.\ several demixed
horizontal layers are formed. However, as can be seen most clearly from the first of the time series of total density profiles along the bottom row of Fig.\ \ref{T05_rho03_a001_x03}, in contrast with the previous cases where
the total density was nearly constant, one can now see that the layers rich in species 2 are of substantially higher density than the ones rich in species 1. Thus, this is initially a gas-liquid type phase separation, where
the liquid-like regions are rich in species 2, while the lower density gas is rich in species 1. As time proceeds, at time $t^*=400$ we see that the horizontal bands across the middle of the system break-up into drops. This,
in turn allows the colloids of species 1 from the neighbouring regions to gather together. The strong inhomogeneity in the density in this region leads to the formation of the third phase, which is a liquid rich in colloids of
species 1. As time proceeds there is also a coarsening as drops of the same phase coalesce to form larger drops. Finally, the system reaches a state consisting of the following microphases: Along the bottom of the system, we
observe a low density (gas-like) layer of variable thickness that is rich in particles of species 1. Above this, there is a liquid layer rich in species 2, above which are several drops composed of the liquid phase rich in
species 1. Next, there are further drops of the liquid phase rich in species 2. All of these liquid drops are surrounded by a gas phase of roughly equimolar composition. Note that because the gravitational driving towards the
bottom of the system is rather weak, the sedimentation of the liquid drops towards the bottom and the displacement of the gas phase towards the top is rather slow (not displayed).

To illustrate the influence of the gravitational force, we now consider the same system (i.e.\ with $T^*=0.5$, $\rho^*=0.3$ and $x=0.3$), but where the external potential driving the colloids towards the bottom of the system
is substantially stronger, with $a_1=a_2=0.1\varepsilon$. In Fig.\ \ref{T05_rho03_a01_x03} we display a time series of concentration profiles and the corresponding total density profiles. As expected, the sedimentation is much
faster now, as can be seen from the total density profiles along the bottom row of Fig.\ \ref{T05_rho03_a01_x03}. In the early stages of the dynamics the strong gravitational driving leads to a large flux of both species of
colloids towards the bottom, leading to a rapid increase in the density in the lower half of the system. This situation may be termed as ``gravity-dominated'', since the external field has a dominating impact on the structure
of the fluid, while the effect of the intermolecular forces is somewhat weaker. However, even during this stage the particles show their tendency to demix, as evident from the time $t^*=100$ concentration plot in the upper
left panel of Fig.\ \ref{T05_rho03_a001_x03}. As time proceeds, the higher density liquid phase formed at the bottom of the system by the strong gravitational driving then undergoes liquid-liquid demixing. This occurs
initially by the system forming alternating horizontal layers of the two different liquid phases. Subsequently, the thinner layers rich in the minority species 1 colloids break-up into drops. This dynamics is driven by the
fact that it lowers the amount of interface between the two liquid phases. The late-time configurations consist of a single layer of the species 2 rich liquid in the lower half of the system, within which are drops of the
species 1 rich liquid (recall that the gravitational force is the same for both species, so neither liquid phase sediments within the other). The gas phase at the top of the system is of a roughly equimolar concentration.

One further thing to note, comparing Fig.\ \ref{T05_rho03_a001_x03} (weak gravity) and Fig.\ \ref{T05_rho03_a01_x03} (strong gravity), is that in the former the wall at the bottom of the system becomes covered by a thin
gas-like layer. Even if disturbed, this thin layer is very stable, since the wall is hard and is thus preferentially adsorbed by the gas phase. However, in the case in Fig.\ \ref{T05_rho03_a01_x03}, the strong gravitational
force overcomes the interfacial effects and pushes the liquid phase onto the wall.

The impact of the gravitational field can also be seen in Fig.\ \ref{y_rho03_T05} where we display plots of the height of the center-of-mass of both species as they vary as a function of time. For all times $t>0$, the
center-of-mass of the majority species has a lower value than that of the minority species. Initially, the center-of-mass of both species decreases as one would expect. However, interestingly, in all cases the center-of-mass
of the minority species is a non-monotonic function of time, decreasing to a minimum value, before increasing again. In the case of weaker external field (upper panel), the initial  ``gravity-dominated'' regime is less
dramatic and the minimum is shifted to later stages. However, if the the time axis is rescaled by a factor proportional to the amplitude of the external field, the positions of the minima are comparable. The center-of mass
rise is especially pronounced when the composition of the mixture is well away from $x=0.5$ and for the weaker field. In this second stage thermodynamic effects become important, so that the liquid-liquid phase separation
prevents the minority component from sedimentating to the bottom.

We now consider the case of an equimolar mixture, i.e.\ with $x=0.5$, sedimenting under the influence of the same strong external field, with $a_1=a_2=0.1\varepsilon$. A time series of local concentration and total density
profiles for this case are displayed in Fig.\ \ref{T05_rho03_a01_x05}. In the same way we saw already in Fig.\ \ref{T05_rho03_a01_x03}, the stronger external field induces a rapid build-up of the liquid above the lower wall,
during the initial ``gravity-dominated'' period of the sedimentation, leading to the formation of a layer of liquid at the bottom of the system, with the gas above. The rapid increase in the density at the bottom of the system
then triggers the liquid-liquid phase separation within the liquid phase. In the bulk fluid, this would result in the formation of two separate large slabs of the two liquid phases, separated by a single interface. However, in
the present case, where the fluid is confined by the walls and also pushed to the bottom by the gravitational force, the system is restricted in its ability to coarsen to form just two slabs, as it would in bulk spinodal
phase-separation. Instead, we see the system break up into vertical stripes alternating between the two different liquid phases, respectively rich in species 1 and species 2. The interfaces between the stripes are vertical
(contact angle of 90$^\circ$ at the lower wall) because the mixture is symmetric and so the wall does not favour either liquid phase over the other. There is some coarsening in the number of stripes, but the system is unable
to reach the minimum free energy state consisting of just one drop of each of the two different liquid phases at the bottom of the system. This is because to reach the minimum the system must surmount a free energy barrier by
a diffusion of particles of one species through a region rich in the other species. Such a process is quite unlikely, given the low temperature of the system. We thus believe that the final time $t^*=600$ plot in Fig.\
\ref{T05_rho03_a01_x05} corresponds to a configuration which is very stable with an extremely long (maybe even infinite) relaxation time for reaching the true equilibrium configuration.

Finally, we now discuss a case where species 1 experiences a stronger driving towards the bottom of the system than species 2. This, for example, corresponds to the case where the species 1 colloids are made of a more dense
material than species 2, and so have a much higher buoyancy mass. In Fig.\ \ref{T05_rho03_a001_a01_x05} we present results for a system with the same parameters as in the previous system, with the only difference that now
$a_2=0.01\varepsilon$, i.e.\ the amplitude of the external field experienced by species 2 is one order of magnitude less than the one experienced by species 1 ($a_1=0.1\varepsilon$). In Fig.\ \ref{T05_rho03_a001_a01_x05} we
see that the much greater sedimentation rate of species 1 leads to a separation of the two species. The center-of-mass of species 2 is almost constant (not displayed). This is because some of the slowly falling species 2 rich
liquid drops are subsequently pushed upwards by the faster sedimenting species 1 rich liquid. In this situation the disparity in the strength of the external driving forces is large enough to overcome any free energy barriers
preventing a drop of one liquid phase being dragged through the other. In the final stages, the system forms a state with the following morphology: a slab of liquid rich in species 1 is located along the bottom of the system,
whilst drops of the liquid phase rich in species 2 remain above this layer with the gas phase which has a roughly equimolar composition intruding between the drops. In contrast with what we observed in Figs.\
\ref{T05_rho03_a01_x03} and \ref{T05_rho03_a01_x05}, these liquid drops are not in close contact with the bottom liquid but instead almost `float' in the gas phase. It is because the external force on them downward is much
weaker and also because the interfacial tension between the liquid phases and the gas phase is less than the interfacial tension between the two liquid phases. Consequently, these drops lower the overall free energy by rising
a short distance above the lower liquid to allow the gas to intrude between the two liquid phases.

\section{Concluding Remarks}\label{sec:IIII}

In this paper, a two-dimensional binary mixture of partially immiscible colloids dispersed in a continuous solvent medium was studied using DDFT. The mixture is
subject to a gravitational field, which pushes the particles down towards the bottom of the system at $y=0$, where there is an impenetrable hard-wall. In the
absence of the gravitational field, the contact angle of each individual species is $180^\circ$ for all temperatures $T<T_c$, as usual for a fluid at a hard
wall. However, we have seen that the presence of the other species and the gravity strongly influences the interfacial properties at the wall. This in turn has a
significant effect on the structure and microphase behaviour of the entire system both in equilibrium and out of equilibrium.

A description of all possible behaviours that our model can exhibit during the sedimentation relaxation towards equilibrium is a lengthy and almost impossible
task. The vast array of possible configurations found in Ref.\ \onlinecite{fathi} using continuation for a mesoscopic model binary fluid on a surface together
with the gas phase above, should also be observed in the present model and gives an indication of what drop morphologies can be observed in the present system.
Our intention here is not to categorise all possible morphologies but instead to point out some of the most characteristic features of the model and to
demonstrate the complexity of its dynamical properties. The systems discussed here differ in temperature, density, composition and/or strength of the
gravitational field. Based on these examples, we have illustrated an interplay between the bulk thermodynamics, interfacial (both solid-fluid and fluid-fluid)
phenomena and gravity that gives rise to a rich variety of behaviours of the system which can often considerably change over time.

The dynamical properties of colloidal suspensions are often modelled using stochastic Brownian dynamics (BD) computer simulations, based on numerically integrating forwards in time the Langevin equations of motion in Eq.\
(\ref{eq:EOM}). Here, we have instead used an alternative microscopic approach, namely DDFT. This has an advantage that for a given size of the system (and a given discretisation) the numerical requirements are independent of
the total fluid density, i.e., systems that correspond to say $\sim10^2$ particles or those corresponding to $\sim10^4$ particles can be handled at the same numerical price. Since BD simulations are too computationally
expensive to implement for the systems considered here, we solely focussed on applying the DDFT. The surprisingly good agreement between the BD simulation results and those from the rather simple DDFT treatment of the model
found previously for the one-component fluid,\cite{archer_mal} gives us confidence that our DDFT approach is also reliable for the binary mixture. Note that we have solely presented results from 2D calculations. Since the
external potential is translationally invariant along the horizontal axis, in principle one need only treat the system as effectively in 1D. However, using an approximate DFT in 1D leads to poor results -- see e.g.\ Fig.\ 9 of
our previous paper dealing with a one-component fluid.\cite{archer_mal} Below we discuss this issue further.

We should stress that here we have assumed that the hydrodynamic interactions between the colloids can be neglected, both in the Langevin equations of motion
(\ref{eq:EOM}) and also in the resulting DDFT (\ref{ddft}). It would be interesting to investigate their influence on the system behaviour, because in certain
situations this can be significant.\cite{Paddy} Hydrodynamic interactions can be included within the DDFT framework -- see e.g.\ Refs.\ \onlinecite{ReLo08} and
\onlinecite{ser}, but we leave this aspect to future work.

Our main observations can be summarised as follows: At high densities and/or high temperatures, the fluid exhibits liquid-liquid demixing but no liquid-gas
transition. In this case, depending on the concentration, we observe three characteristic scenarios:

\begin{enumerate}
\item
When the concentration of one of the species is low, we observe the formation of liquid droplets rich in the minority species that are dispersed within a slab of the majority species  rich liquid. Such a state is preceded by
the formation of horizontal layers that break up rather quickly to form the drops. However, the presence of a large number of small drops is thermodynamically unfavourable due to the large free energy cost of all the
interfaces. The drops thus have a tendency to coalesce whenever two or more of them approach each other. Alternatively, they may also coarsen via the Ostwald ripening mechanism, but the low miscibility of the mixture can lead
to this microemulsion-like density distribution being stable over a very long time.
\item
At a somewhat higher concentrations of the minority species (but still for $x\neq0.5$), horizontal stripes also form in the initial stages, but in this situation some of the stripes persist without breaking-up.
\item
For the equimolar $x=0.5$ case, the morphology of the fluid is striking different. This is due to the symmetry of the mixture and so the confining walls do not direct the spinodal decomposition. The phase separation is
initiated in the bulk and leads to vertical rather than horizontal stripes, since the final equilibrium state must be invariant with respect to exchanging the labels on the two different species (i.e.\ the possibility of a
preferential adsorption of either of the two species is excluded by symmetry).
\end{enumerate}

We have also examined the behaviour of the mixture at a state point  below the critical point of the liquid-gas transition. In this case, the system separates into three phases: a gas and two liquid phases. To the best of our
knowledge, what we present here is the first microscopic study of the dynamics of such a system. We focus in particular on the influence of gravity, displaying results for three examples: 1) the gravitational field is weak and
same for both components, 2) the gravitational field is strong and same for both components, and 3) the gravitational field is strong for species 1 but weak for species 2. All cases exhibit a very rich microstructure, not only
due to the presence of the three phases but also in the manner the different phases are formed and distributed over different parts of the system, forming different shapes and also in the manner they change over time.
Coalescence of regions of one phase is also hindered by the presence of the other phases. The effect of the gravitational driving can be profound, with the center-of-mass of the minority species being a non-monotonic function
of time, first decreasing and then, surprisingly, moving up again due to the capillary forces in the system.

We conclude with a few general remarks on the nature of the results obtained from a 2D implementation of DDFT -- see also the discussion in Ref.\
\onlinecite{archer_mal}. In the limit $t\to\infty$, the resulting configuration corresponds to the density profiles for an equilibrium fluid, as obtained from
DFT, since from Eq.\  (\ref{ddft}), at equilibrium we have \bb \frac{\delta F[\{\rho_i\}]}{\delta \rho_i(\rr)}=const. \label{EL_eq} \ee Such equilibrium profiles
must, of course, reflect the symmetry of the external potentials since DFT is a statistical mechanical theory, which in principle performs a statistical average
over all realisations of the ensemble. This condition should be obeyed no matter what approximation is made for the free energy functional or whether the
solution is obtained from a 1D or a 2D calculation. Our results, where we find density profiles that vary in the $x$-direction, the direction in which the
external potentials are invariant, lead to the obvious question: Why do we observe this symmetry breaking? This is a particularly pertinent question, since the
results we find do agree qualitatively with what one would observe in any individual realisation of the ensemble -- see e.g.\ the comparisons for the
one-component fluid in Ref.\ \onlinecite{archer_mal}.

To address this question, it is important to stress that Eq.\ (\ref{ddft}) is a {\emph{deterministic}} theory for a local and instantaneous density distribution, i.e. for the local density averaged over the noise realization.
Clearly then, if both the external field and the initial density distribution only vary in the vertical direction, then $\rho(x,y,t)$ can also change only in that direction and the solution of the 2D problem is exactly the
same as if the system is treated as a 1D problem. Nevertheless, the perturbation field $\delta_\rho(x,y)$ added to the otherwise uniform initial density profile may break the symmetry and allows in principle for density
profiles that exhibit inhomogeneities even along the $x$-axis. If the system contains a thermodynamically stable phase, then these density fluctuations quickly disappear. However, inside the spinodal (as in our case), the
fluid is thermodynamically unstable and it follows that there exist some density fluctuation modes whose amplitude will grow exponentially with time (at least at early times). The length scale corresponding to the fastest
growing mode can be calculated for the bulk fluid from the dispersion relation\cite{archer,ArEv04} resulting from linearizing Eq.\ (\ref{ddft}). This length scale is typically $10\sigma$--$20\sigma$, depending on the state
point and is in good agreement with the length scale observed at early times in our density profiles.

Another important point to be emphasized is that DDFT is a theory that is formulated in the canonical ensemble, with a fixed number of particles $N=N_1+N_2$. In contrast, DFT is formulated in the grand canonical ensemble,
which is coupled to a reservoir that fixes the chemical potentials $\mu_i$ and with which the system can exchange particles. In the limit $t\to\infty$ DDFT reduces to the Euler-Lagrange equation (\ref{EL_eq}) subject to the
additional condition $\int\dr (\rho_1(\rr)+\rho_2(\rr))=N$. Thus, even in the equilibrium limit, the DFT and DDFT results do not generally coincide, unless a proper thermodynamic limit (i.e., $N\to\infty$) is taken. For any
finite values of $N$ (and volume $V$), though, the finite size of the system acts as an external constraint, which plays a role somewhat akin to a stabilizing field, for interfacial structures that are metastable in the
thermodynamic limit.\cite{Hend} A well-known example is a nucleating drop of liquid in a supersaturated vapor.\cite{mal_jack}

All these points hold irrespective of whether the free energy functional used in the theory is exact or approximate. Clearly, for any physically relevant system one always
deals with an approximate functional and as shown in our previous work for the one-component fluid,\cite{archer_mal} $\rho(y)$ is more accurate when obtained from
2D DFT after averaging $\rho(x,y)$ over the $x$-direction (the direction parallel to the wall) rather than directly from the 1D DFT. We believe this is due to
the fact that within our mean-field treatment some of the fluctuation contributions to the free energy, such as capillary waves, are inevitably ignored. This
issue is also nicely discussed by Reguerra and Reiss,\cite{reguerra} who studied what fluctuation effects are underestimated in the standard approximations used
in DFT when used to study drops of liquid confined in a cavity.
As discussed in Ref.~\onlinecite{archer_mal}, within 2D (D)DFT one relaxes some of the constraints made in a 1D (D)DFT calculation, and thus the resulting averages are made over a larger set of microstates.

This situation is akin to the following simple mechanical example: Consider a cylindrical elastic rod with its tip in contact with the floor and its axis of symmetry perpendicular to the floor.  If the rod is gently pressed
down on the opposite tip with a force parallel to the axis of symmetry, then the rod will bend slightly outwards in the middle, breaking the symmetry. The direction in which it bends is arbitrary. If the experiment is repeated
numerous times and an average over all the positions of the rod is calculated, then the average state will consist of an unbent rod, since all the different realisations compensate one another. Thus, such a series of
experiments will give the same ``mean'' of the rod shape as a ``1D'' consideration but contains more information about the shape fluctuations of the rod. Accounting for the states corresponding to when the rod is bent in a
particular direction is, in some sense, the way the present system is treated by the approximate 2D (D)DFT.

\begin{acknowledgments}
 \noindent A.M. acknowledges support from the Czech Science Foundation, project 13-09914S.
\end{acknowledgments}

\end{document}